\documentclass[conference]{IEEEtran}
\IEEEoverridecommandlockouts

\usepackage{algorithmic}
\usepackage[linesnumbered,ruled,vlined]{algorithm2e}
\usepackage{multirow}
\usepackage{multicol}
\usepackage{wrapfig}
\usepackage{soul}
\usepackage{balance}

\usepackage{caption}
\usepackage{subcaption}
\usepackage{graphicx} 
\usepackage{subcaption} 
\usepackage{subfig}

\usepackage{epstopdf}

\usepackage{amsfonts}
\usepackage[flushleft]{threeparttable}
\usepackage{adjustbox}
\usepackage{fancyhdr}
\usepackage[dvipsnames]{xcolor}
\usepackage[flushleft]{threeparttable}
\usepackage{tikz}
\usepackage{outlines}
\usepackage{tcolorbox}
\usepackage{blindtext}
\usepackage{xcolor}
\usepackage{color}
\usepackage{listings}
\usepackage{color}
\usepackage{MnSymbol,wasysym}
\usepackage[switch]{lineno}
\usepackage{array}

\usepackage{amsmath,amsthm}

\usepackage{mathtools} 

\usepackage[mathscr]{euscript}
\usepackage{caption} 

\usepackage{xcolor,colortbl}
\def\BibTeX{{\rm B\kern-.05em{\sc i\kern-.025em b}\kern-.08em
    T\kern-.1667em\lower.7ex\hbox{E}\kern-.125emX}}

\usepackage{tikz}

\usepackage[T1]{fontenc} 

\usepackage{subcaption}

\theoremstyle{plain}

\begin{document}

\title{Predictive Performance of Photonic SRAM-based In-Memory Computing for Tensor Decomposition}

\author{
    \IEEEauthorblockN{Sasindu Wijeratne\IEEEauthorrefmark{2}, Sugeet Sunder\IEEEauthorrefmark{1}, Md Abdullah-Al Kaiser\IEEEauthorrefmark{3}, Akhilesh Jaiswal\IEEEauthorrefmark{3} \\ Clynn Mathew\IEEEauthorrefmark{1}, Ajey P. Jacob\IEEEauthorrefmark{1}, Viktor Prasanna\IEEEauthorrefmark{2}}
    \IEEEauthorblockA{\IEEEauthorrefmark{2}Ming Hsieh Department of Electrical and Computer Engineering, University of Southern California}
    \IEEEauthorblockA{\IEEEauthorrefmark{1}Information Sciences Institute (ISI), University of Southern California}
    \IEEEauthorblockA{\IEEEauthorrefmark{3}Electrical and Computer Engineering, University of Wisconsin-Madison}
    Email: \{kangaram, prasanna\}@usc.edu, \{sunder, cmathew, ajey\}@isi.edu, \{mkaiser8, akhilesh.jaiswal\}@wisc.edu
}


\maketitle
\begin{abstract}
Photonics-based in-memory computing systems have demonstrated a significant speedup over traditional transistor-based systems because of their ultra-fast operating frequencies and high data bandwidths. Photonic static random access memory (pSRAM) is a crucial component for achieving the objective of ultra-fast photonic in-memory computing systems. In this work, we model and evaluate the performance of a novel photonic SRAM array architecture in development. Additionally, we examine hyperspectral operation through wavelength division multiplexing (WDM) to enhance the throughput of the pSRAM array. We map Matricized Tensor Times Khatri-Rao Product (MTTKRP), a computational kernel commonly used in tensor decomposition, to the proposed pSRAM array architecture. We also develop a predictive performance model to estimate the sustained performance of different configurations of the pSRAM array. Using the predictive performance model, we demonstrate that the pSRAM array achieves 17 PetaOps while performing MTTKRP in a practical hardware configuration.

\end{abstract}

\begin{IEEEkeywords}
Photonic Computing, MTTKRP, Tensor Decomposition
\vspace{-3mm}
\end{IEEEkeywords}

\section{Introduction}
Recent advancements in analyzing large datasets have led to information being inherently represented as higher-order data structures known as tensors. Tensor decomposition converts input tensors into a reduced latent space, which can then be utilized to identify important features of the underlying data distribution. Tensor decomposition has been effectively used in various fields, such as machine learning, signal processing, and network analysis~\cite{mondelli2019connection,cheng2020novel,wen2020tensor}. Additionally, tensor decomposition has been instrumental in improving the interpretability of complex models by breaking down multi-dimensional data into simpler, more manageable components. This decomposition technique has also facilitated advancements in areas such as bioinformatics~\cite{taguchi2020tensor}, where it aids in the analysis of multi-modal biological data, and in computer vision~\cite{panagakis2021tensor}, where it enhances image and video processing tasks. The flexibility and robustness of tensor decomposition methods continue to drive innovation across a wide range of scientific and engineering disciplines. 

Canonical Polyadic Decomposition (CPD) is arguably the most widely used method to decompose a tensor into a low-rank tensor decomposition model~\cite{kolda2009tensor, aggour2019intelligent}. It has become the standard tool for unsupervised multi-way data analysis. The Matricized Tensor Times Khatri-Rao Product (MTTKRP)~\cite{8821030} is recognized as the most time-consuming computational kernel in CPD. Due to the irregular shapes of the real-world tensors, specialized hardware accelerators are increasingly popular to enhance the efficiency of sparse tensor computations~\cite{10.1145/3543622.3573179, wijeratne2021programmable, 9622851,9065579}.

In recent years, digital electronics have significantly advanced the power and performance metrics of digital computing systems~\cite{khan2018science}. The 6-transistor electrical SRAM has become the standard for on-chip memory storage. However, these electrical SRAMs fall short in terms of computing speed, throughput, and power efficiency for new data-intensive applications such as machine learning, signal processing, and large-scale simulations~\cite{7891546}. The rapid increase in the energy delay product associated with data movement and memory access for compute operations underscores the memory bottleneck that affects modern digital computing systems~\cite{khan2018science}. Solutions involving parallel, near-memory, and in-memory processors based on electronic physical-state variables do not completely address the issue due to constraints related to scalar computing and the latency of long metal interconnections.

Recently, photonic SRAMs have emerged as a promising alternative to electronic charge-based SRAMs~\cite{osram_arxiv}. Numerous photonic SRAM implementations have been investigated in the past \cite{pleros2008optical, tsakyridis201910, dong2015nano, li2009optical, alexoudi2016iii, pitris2016wdm, liu2006packaged, trita2013monolithic}. However, developing a photonic SRAM technology that is compatible with current foundry manufacturing processes and offers ultra-high speed and low energy consumption remains a significant challenge.

In this work, we introduce a novel photonic SRAM array embedded in a scalable optical in-memory compute engine, designed using existing foundry processes. The photonic SRAM uses available GF45SPCLO photodiodes and ring resonators, which are upgrades over our previous designs~\cite{osram_arxiv} and utilized models of photonic devices reported in the literature~\cite{jacob2024non,jacob2024electro}. Our approach combines the high-speed and bandwidth advantages of photonic technology with the proven reliability of SRAM while addressing the challenges of integrating optical components with standard CMOS processes to create a scalable and efficient in-memory computing solution. 



The contributions of our paper are as follows:
\begin{itemize}
\item A novel embedded photonic SRAM (pSRAM) array is designed and implemented in a scalable optical in-memory computing engine, operating in the O-band. The pSRAM is reconfigurable at speeds exceeding 20 GHz; the write speed of the RAM. The overall performance of the pSRAM array is determined by the speed of the optical components that constitute the system architecture.

\item We map the compute primitives of MTTKRP to the pSRAM array architecture.
\item We develop a predictive performance model to evaluate the sustained performance of the proposed photonic SRAM memory architecture on MTTKRP.
\item The predictive performance model shows that the pSRAM array can achieve a sustained performance of 17 PetaOps with 8-bit precision under practical pSRAM array configuration.
\end{itemize}

\section{Background}~\label{background}

A tensor is a generalization of an array in multiple dimensions. In TD, the number of dimensions of an input tensor is commonly called the number of tensor modes. A real-valued $N$ mode tensor is denoted by $\mathcal{X} \in \mathbb{R}^{I_0 \times \cdots \times I_{N-1}}$. Further, $\mathcal{X}_{(n)}$ denotes the mode-$n$ matricization or the unfolding~\cite{favier2014overview} of the matrix $\mathcal{X}$. $\mathcal{X}_{(n)}$ is defined as the matrix $\mathcal{X}_{(n)} \in \mathbb{R}^{I_n \times (I_0 \cdots I_{n-1} I_{n+1} \cdots I_{N-1})}$ where parenthetical ordering indicates that the mode-$n$ column vectors are arranged by sweeping all the other mode indices through their ranges.

Canonical Polyadic Decomposition (CPD) decomposes $\mathcal{X}$ into a sum of single-mode tensors (i.e., arrays), which best approximates $\mathcal{X}$. For example, given 3-mode tensor $\mathcal{X} \in \mathbb{R}^{I_0 \times I_1 \times I_2}$, our goal is to approximate the original tensor as
\vspace{-3mm}
\begin{equation} \label{eqn_approx_tensor}
\begin{split}
\mathcal{X} \approx \sum_{r=0}^{R-1} \mathbf{a}_r \circ \mathbf{b}_r \circ \mathbf{c}_r
\end{split}
\end{equation}
\vspace{-3mm}

where $R$ is a positive integer and $\mathbf{a}_r \in \mathbb{R}^{I_0}$, $\mathbf{b}_r \in \mathbb{R}^{I_1}$, and $\mathbf{c}_r \in \mathbb{R}^{I_2}$. For a thorough review of CPD, refer to~\cite{kolda2009tensor}.

In the rest of Section~\ref{background}, we assume that the number of modes is three for illustration purposes.


\vspace{-3mm}
\begin{algorithm}
\DontPrintSemicolon
Input: A tensor $\mathcal{X} \in \mathbb{R}^{I_0 \times I_1 \times I_2}$, the rank $R \in \mathbb{Z}^{+}$ \;
Output: CP decomposition $[\![ \mathbf{A}, \mathbf{B}, \mathbf{C} ]\!]$, $\mathbf{A} \in \mathbb{R}^{I_0 \times R}$, $\mathbf{B} \in \mathbb{R}^{I_1 \times R}$, $\mathbf{C} \in \mathbb{R}^{I_2 \times R}$ \;
\While{\emph{stopping criterion not met}}{
\textcolor{blue}{// Matricization of $\mathcal{X}$ is different for each factor matrix computation} \;

    $\mathbf{A} \gets \mathbf{spMTTKRP}(\mathcal{X}_{(0)}, \mathbf{B}, \mathbf{C})$ \;
    $\mathbf{B} \gets \mathbf{spMTTKRP}(\mathcal{X}_{(1)}, \mathbf{A}, \mathbf{C})$ \;
    $\mathbf{C} \gets \mathbf{spMTTKRP}(\mathcal{X}_{(2)}, \mathbf{A}, \mathbf{B})$ \;
   Normalize $\mathbf{A}$, $\mathbf{B}$, $\mathbf{C}$ \;
}
\caption{CP-ALS for a 3-mode tensor}
\label{cp-als}
\end{algorithm}

The alternating least squares (ALS) method is used to compute CPD. Algorithm~\ref{cp-als} shows the ALS method for CPD (i.e., CP-ALS) where Matricized Tensor-Times Khatri-Rao product (MTTKRP) is iteratively performed on all the Matricizations of $\mathcal{X}$, iteratively. In this paper, performing MTTKRP on all the Matricizations of an input tensor is called computing MTTKRP along all the modes. The outputs $\mathbf{A}$, $\mathbf{B}$, and $\mathbf{C}$ are the factor matrices that approximate $\mathcal{X}$. $\mathbf{a}_r$, $\mathbf{b}_r$, and $\mathbf{c}_r$ in Equation~\ref{eqn_approx_tensor} refers to the $r^{\text{th}}$ column of $\mathbf{A}$, $\mathbf{B}$, and $\mathbf{C}$, respectively. 

\section{Optical SRAM Architecture}\label{optical_sram}

\subsection{Input Encoding:}
One significant advantage of employing photonic devices in computing architectures is leveraging hyperspectral computing using wavelength division multiplexing (WDM)~\cite{xu2022neuromorphic}.
WDM allows a single optical channel to carry multiple data streams by simultaneously transmitting signals at different wavelengths, maximizing data transmission capacity without interference. Optical frequency combs, generated using microresonators, produce a precise series of narrow spectral lines (comb lines) spanning a broad range of wavelengths~\cite{chang2022integrated}.

The proposed device operates in the O-band, offering 52 wavelength channels (based on Global Foundries 45SPCLO PDK) with sub-nanometer spacing for efficient data transmission. 
In this system, we envision an intensity encoded input data, with each discrete power level corresponding to a specific value represented by an 8-bit word. To modulate multiple wavelength channels simultaneously with varying intensity levels, we employ comb shapers—optical devices designed to manipulate the spectral properties of an optical frequency comb. High-speed Electro-optic modulators are a common type of comb shapers that selectively attenuate or enhance specific comb lines, allowing for precise shaping of the comb spectrum for various applications. 


\subsection{Bitcell:}

\begin{figure}[ht]
     \centering
     \begin{subfigure}[b]{0.49\textwidth}
         \centering
         \includegraphics[width=\textwidth]{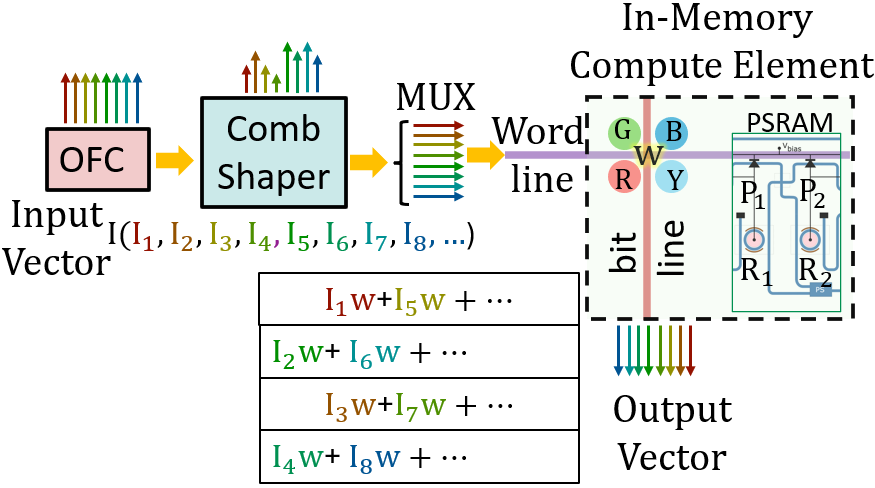}
         \caption{}
     \end{subfigure}
     
     \vspace{4mm}
     
     \begin{subfigure}[b]{0.35\textwidth}
         \centering
         \includegraphics[width=\textwidth]{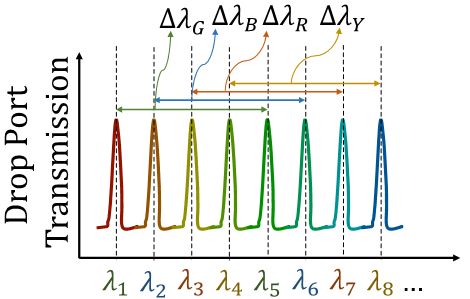}
         \caption{}
     \end{subfigure}
        \caption{(i) Schematic of the proposed computing engine. Optical frequency combs (OFC) are used to generate precise wavelength channels, which are then modulated using high-speed comb-shapers. The input, encoded across multiple independent wavelength channels, is sent into the word-line and multiplied with a memory bit. Different ring modulators (G/B/R/Y) are employed to handle different sets of wavelengths, with the resonances of other three resonators spaced within the FSR of the one. An analog output is received on the bit-line for further processing. (ii) The drop port transmission characteristics of the compute ring modulators indicates the spacing of wavelength channels used for WDM.}
       \label{fig:bitcell}
\end{figure}

Conventional electrical SRAMs face significant speed and power consumption bottlenecks due to the large bitline/ wordline capacitance and high interconnect resistance due to technology scaling \cite{sram_bottleneck}. In contrast, our proposed photonic SRAM (pSRAM) exhibits ultra-low energy consumption and high-speed read/write operation \cite{osram_arxiv}. To construct the optical latch structure, the pSRAM bitcell employs cross-coupled microring resonators (MRR) and photodiodes (PD) illustrated in Figure~\ref{fig:bitcell}. The through port of MRR R1 (R2) drives the photodiode P2 (P1), which controls the resonance state of the other MRR R2 (R1). Hence, the cross-coupled structure ensures the storing of differential optical data inside the latch. 
The pSRAM bitcell is projected to operate at a 20 GHz frequency while consuming $\sim$1.04 pJ/bit ($\sim$16.7 aJ/bit) switching (static) energy~\cite{osram_arxiv}.
The proposed pSRAM is designed as a 2D crossbar array of memory bitcells, with each cell connected to a pair of bit lines and a word line. Typically, a word line is one word and controls the activation of read/write operations for the corresponding bitcells. While the read speed of pSRAM is faster, constrained by the time constant of ring resonators. The write speed of pSRAM is currently at 20 GHz, which determines the reconfigurability rate of pSRAM.


\color{black} Furthermore, the fabrication-friendly architecture of our pSRAM subsystem, which is based on a pSRAM prototype designed on GF 45SPCLO PDK, has been sent to the fab for tapeout, enabling seamless integration alongside the electrical subsystem.

\subsection{Output Encoding:}

As bitcell can only store binary data, appropriate intensity scaling depending on the bit-position is required for the compute operation. The dot product of the 8-bit intensity encoded input and a 8-bit binary word stored in the pSRAM array results in 8 analog optical outputs. Each optical output obtained is inherently scaled according to its corresponding bit significance, with the maximum optical power delivered to the bit representing the most significant digit (MSB) and appropriately scaled power sent into the least significant bit (LSB). These optical outputs are converted and accumulated through the photocurrents of the photodetectors. The analog accumulated photocurrent values can be converted into digital electrical bitstreams through high-speed on-chip analog-to-digital converters (ADC).
Integrating on-chip ADC facilitates the seamless conversion of analog optical signals to digital electrical form, which enables on-chip CMOS hardware/accelerator for further processing in the electrical domain. 



\section{Mapping MTTKRP Computational Primitives to pSRAM Array} 

\subsection{Grid Representation of pSRAM Array} 

\begin{figure}[ht]
\vspace{-2mm}
\centering
\includegraphics[width=0.7\linewidth]{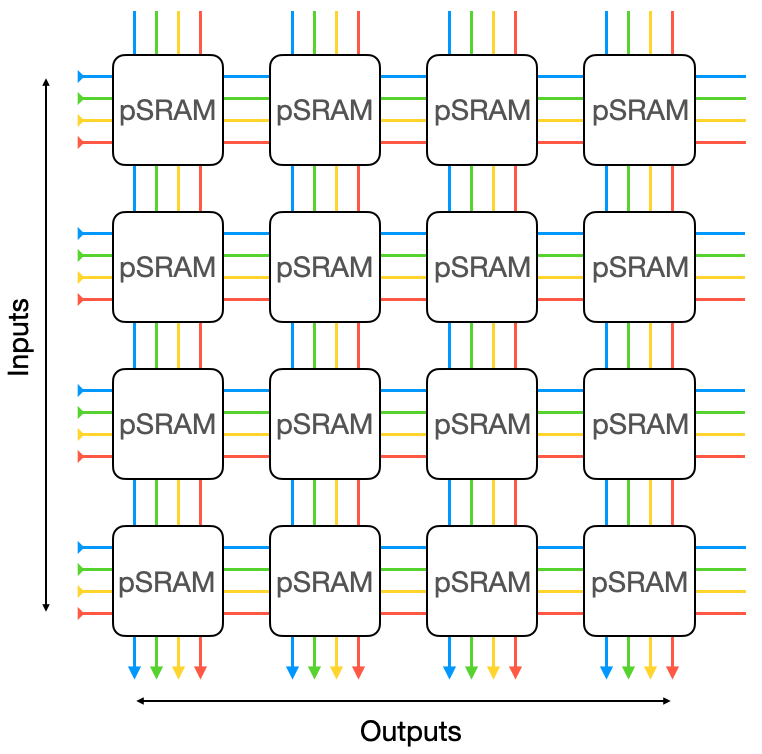}
\vspace{-2mm}
\caption{Grid representation of pSRAM array.}
\label{fig:grid_rep}
\end{figure}

Figure~\ref{fig:grid_rep} illustrates the 2D grid representation of the pSRAM array. In this figure, each pSRAM word (group of pSRAM cells) is shown as a square, while a wavelength is shown as a line. Since the design supports hyperspectral encoding, different color lines in Figure~\ref{fig:grid_rep} represent different wavelengths. For demonstration purposes, Figure~\ref{fig:grid_rep} displays a 2D grid containing $4 \times 4$ pSRAM words and 4 distinct wavelengths. Each pSRAM is capable of multiplying the values stored within the word by the inputs from the wavelengths. The addition is conducted along each column by summing the intensity of identical wavelengths.

\begin{figure}[ht]
\centering
\includegraphics[width=0.8\linewidth]{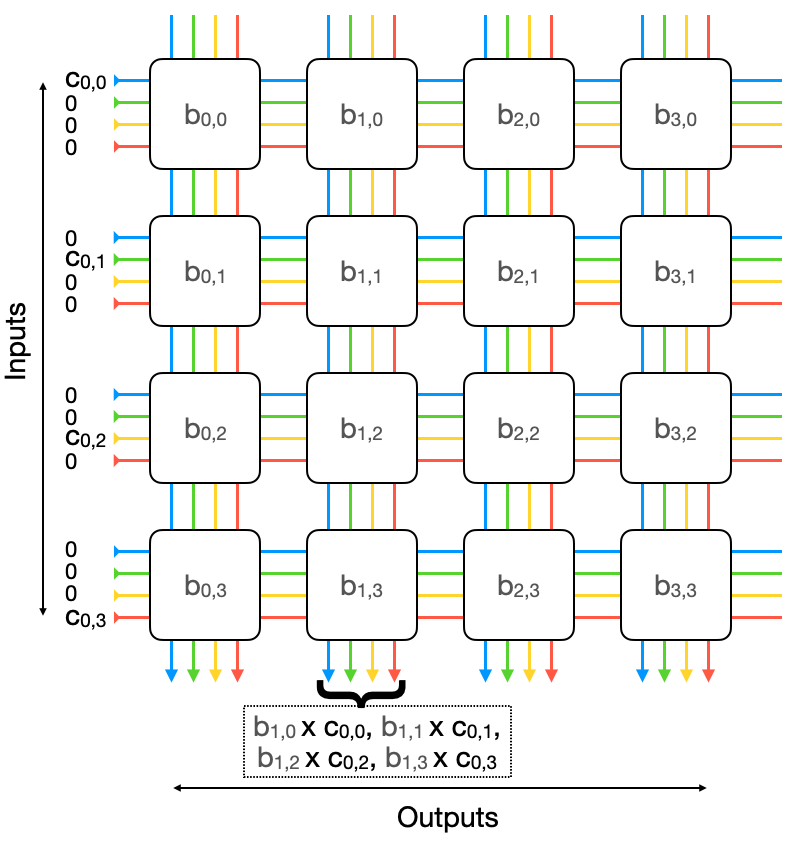}
\vspace{-2mm}
\caption{Mapping CP 1 to pSRAM array.}
\label{fig:hadamard_product_isi}
\vspace{-6mm}
\end{figure}

\subsection{Computational Primitives}

We identified 3 computational primitives (CPs) that the pSRAM array should support to perform MTTKRP.


In this Section, we describe the computational primitives using a 3-mode tensor $\mathcal{X}$ and its factor matrices $\textbf{A}$, $\textbf{B}$, and $\textbf{C}$. We consider an example in which MTTKRP is performed on the tensor $\mathcal{X}$ to generate the factor matrix $\textbf{A}$. Note that the computational primitives introduced in this Section can be extended to any tensor with any number of modes.

\subsection{Hadamard Product of Factor Matrix Rows (CP 1)} This primitive involves computing the Hadamard Product, which is an elementwise multiplication of corresponding elements of two vectors. For example, given the rows of the factor matrix \( b_{j} \) and \( c_{k} \), the Hadamard product is represented as \( b_{j} \circ c_{k} \).

As illustrated in Figure~\ref{fig:hadamard_product_isi}, a row $\mathbf{b}_i$ (i.e., $i = 0, 1,2 ...$) of the factor matrix $\mathbf{B}$ is loaded and stored in each column of the pSRAM array. Subsequently, each row of the factor matrix $\mathbf{C}$ (denoted as $\mathbf{c}_i$) is loaded, and each element in $\mathbf{c}_i$ is multiplied by the corresponding element in $\mathbf{b}_i$. We interleave the wavelengths while feeding the inputs to avoid addition among column values, as shown in Figure~\ref{fig:hadamard_product_isi}. The resulting product from each column of the pSRAM array is the Hadamard Product of the respective rows of the factor matrices $\textbf{B}$ and $\textbf{C}$. Figure~\ref{fig:hadamard_product_isi} shows only the output from a single column of the pSRAM array. Note that all columns in the grid generate outputs simultaneously.

\begin{figure}[ht]
\centering
\includegraphics[width=0.8\linewidth]{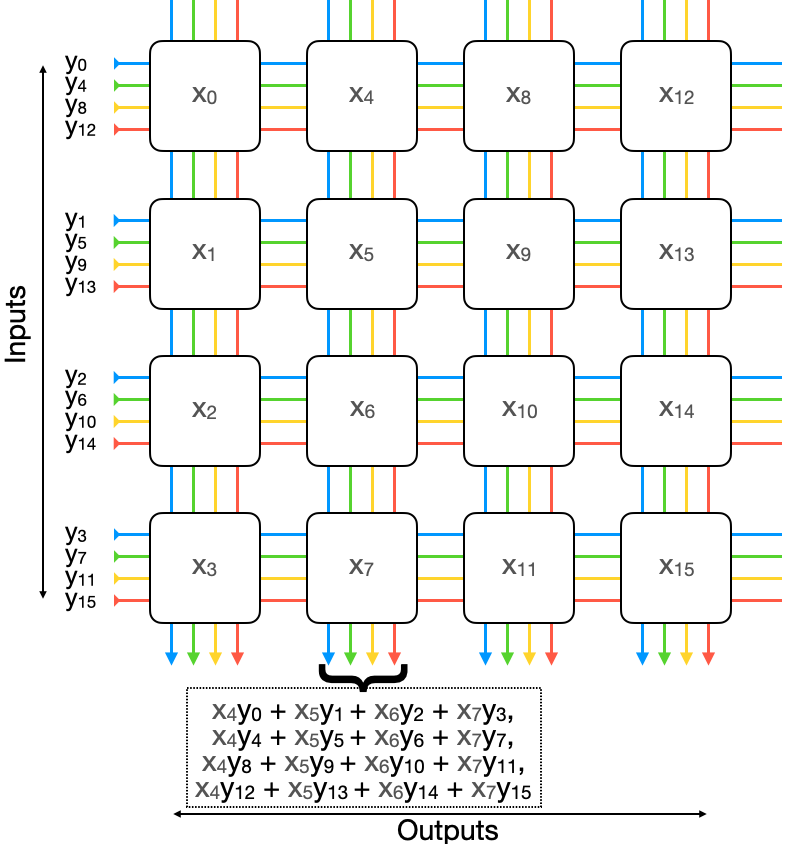}
\caption{Mapping CP2 and CP3 to pSRAM array.}
\label{fig:cp2_cp3_isi}
\vspace{-6mm}
\end{figure}

\subsection{Scaling with a Tensor Element (CP 2)} Using the Hadamard Product (results from CP 1), the second computational primitive multiplies the resulting vector by the respective tensor element \( x_i \). Formally, this can be written as \( x_i \cdot (B_{j_0} \circ C_{k_0}) \).

\subsection{Elementwise Vector Addition to Generate the Factor Matrix (CP 3)}
The third primitive adds the scaled vectors produced in CP 2 to the corresponding row of the factor matrix, $\textbf{A}$ through vector addition. This operation is given by \( A_{i_0} + x \cdot (B_{j_0} \circ C_{k_0}) \), where \( A_{i_0} \) is a row from the factor matrix \( A \).

As shown in Figure~\ref{fig:cp2_cp3_isi}, CP 2 and CP 3 are mapped to pSRAM to produce the final output of \( A_{i_0} + x_i \cdot (B_{j_0} \circ C_{k_0}) \). Tensor elements (indicated as $x_i$) are loaded and stored inside the pSRAM words, and Hadamard products of the matrix columns of factor matrices (shown as $y_i$) are loaded as input using different wavelengths. Similar to Figure~\ref{fig:hadamard_product_isi}, a single output of a column of the pSRAM array is shown in Figure~\ref{fig:cp2_cp3_isi}.

\section{Evaluation}~\label{experiments} \vspace{-6mm}
\subsection{Experiments Setup} \label{ex_setup}

\begin{figure}[ht]
     \centering
     \begin{subfigure}[b]{0.38\textwidth}
         \centering
         \includegraphics[width=\textwidth]{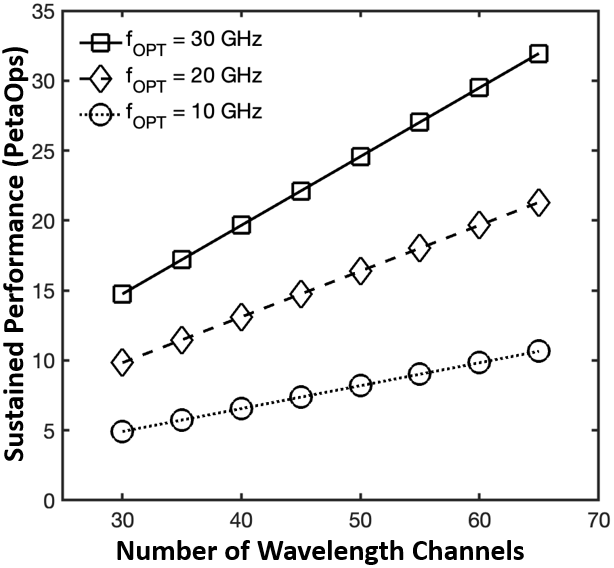}
         \caption{}
         \label{wavelength_channels}
     \end{subfigure}
     
     \vspace{6mm}
     
     \begin{subfigure}[b]{0.38\textwidth}
         \centering
         \includegraphics[width=\textwidth]{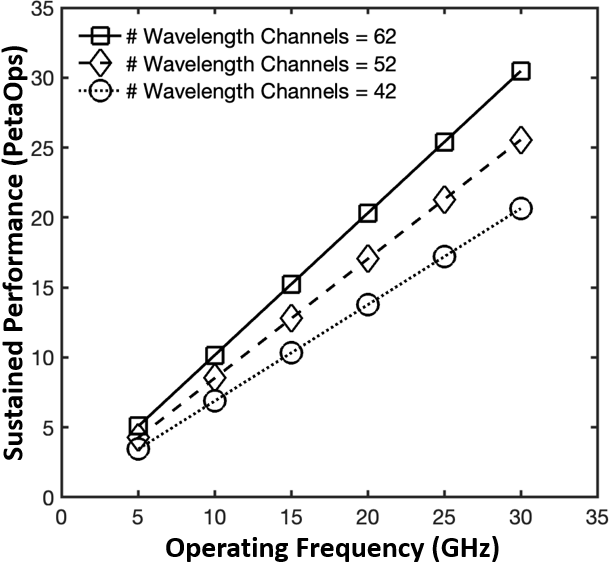}
         \caption{}
         \label{operating_frequency}
     \end{subfigure}
        \caption{(i) Impact of wavelength channels. (ii) Impact of operating frequency.}
       \label{fig:perf_compare_isi}
\end{figure}

The proposed pSRAM array has 256$\times$256 bits. In each row of the pSRAM array, 8 bits are collected together as a word to support 8-bit precision, creating an array of 256$\times$32 words.

Using performance modeling, we evaluated the sustained performance of the pSRAM array while executing MTTKRP on very large tensors (e.g., a 3-mode dense tensor with 1 million indices in each mode).

\subsection{Overall Performance} \label{perf}

Through hardware simulations and performance modeling, we identified that the operating frequency and the number of wavelength channels are the most critical hardware parameters that impact the sustained performance of the proposed architecture. As shown in Figure~\ref{fig:perf_compare_isi}, the sustained performance of MTTKRP on the proposed pSRAM array linearly increases as the operating frequency and the number of wavelength channels increase. Our initial hardware implementation and simulation results show that the proposed pSRAM array can support 52 wavelength channels while operating at 20 GHz. Under these conditions, the proposed optical array achieves a sustained performance of 17 PetaOps while performing MTTKRP.

\section{Conclusion}~\label{conclusion}
\vspace{-4mm}

In this work, we evaluate the performance of a novel photonic SRAM in-memory compute array architecture employing a predictive performance model. We mapped the compute primitives of MTTKRP to the pSRAM array architecture and demonstrated that the architecture can achieve 17 PetaOps in a practical hardware configuration. This work demonstrates the usefulness of photonic in-memory scalar and hyperspectral computing systems that can accelerate complex data-intensive tasks such as MTTKRP.



\section*{ACKNOWLEDGEMENT}
This work is supported by DARPA under grant N660012424003 and the National Science Foundation (NSF) under grant CNS-2009057.
\bibliographystyle{IEEEtran}
\bibliography{reference}

\end{document}